\documentclass[aps,prl,twocolumn,groupedaddress]{revtex4}

\usepackage{graphicx}
\usepackage{dcolumn}
\usepackage{bm}

\begin{document}

\title{Ising Spin Glasses : interaction distribution dependence of the critical exponents}
\author{P. O. Mari}
\affiliation{Laboratoire de Physique des Solides,
\\ Universit\'e Paris Sud, 91405 Orsay, France\\}
\author{I.A. Campbell}
\affiliation{Laboratoire des Verres, \\ Universit\'e Montpellier II,
       34095 Montpellier, France}

\date{\today}

\begin{abstract}
Continuous phase transitions are catalogued into universality classes, families of systems having identical values of all the exponents governing the critical behaviour of their different physical properties. Numerical simulations have been carried out on Ising Spin Glasses in dimension three, using a technique where corrections to finite size scaling can be controled. The data show that the critical exponents vary strongly as a function of  the kurtosis of the interaction distribution, a parameter which from the standard point of view should not be pertinent. This observation implies that for spin glasses the renormalization group analysis should not be approached in a the same way as in the case of canonical second order transitions; a much richer structure of universality classes would appear to exist for spin glasses.
\end{abstract}


\maketitle

The renormalisation group (RG) provides an explanation of the physical origin of the critical exponents and of the universality classes for standard second order transitions which is one of the most remarkable achievements of statistical physics.
The university rules state that the critical exponents depend only on a small number of basic parameters, typically the dimension of space $d$ and the number of order parameter components $n$ \cite{ma}.
Close to a transition, large fluctuating correlated clusters form; universality can be viewed as the consequence of the large scale fractal structure of these clusters being independent  of the details at the microscopic level of the interactions between the elements (spins, atoms,...) making up the system. The few known exceptions to universality concern mostly rare but well understood marginal cases, such as certain regularly frustrated spin systems in two dimensions, where critical exponents vary continuously with the value of a control parameter.

Glass transitions remain enigmatic despite a considerable theoretical and experimental effort \cite{debenedetti}; even the existence or otherwise of a thermodynamic transition in structural glasses is still the subject of intense debate. Spin Glasses can be considered the magnetic analogues of the familiar structural glasses; the canonical examples are made up of local magnetic moments which interact randomly with each other. For technical and conceptual reasons the spin glass transition is much easier to study than structural glass freezing; in particular powerful numerical techniques can be readily applied for Ising Spin Glasses (ISGs). Although spin glass freezing is an unorthodox form of transition as compared with standard second order transitions, extensive experimental and numerical work has demonstrated that these freezing transitions show true continuous transition critical behaviour albeit with exponents which are very different from standard second order transition values \cite{monod,vincent,ogielski,bhatt}. Determining transition temperatures and critical exponents to high precision at spin glass transitions, either experimentally or through simulations, is however a notoriously difficult task. We estimate values for the freezing temperatures and critical exponents of ISGs in dimension three using numerical techniques where  systematic errors can be kept carefully under control. Confirming earlier work \cite{bernardi,mari}, we find sets of critical exponent values which vary strongly with the form of the interaction distribution, in contradiction to the conventional universality rules.

Problems that must be faced for numerical simulations in ISGs include the agonisingly slow equilibration rates, and the need to average over numerous microscopically inequivalent samples to obtain a result representative of the true mean over all possible samples with a given random interaction rule.  Calculations have necessarily been restricted to moderate size samples, finite size scaling rules being used to evaluate the thermodynamic limit ordering temperature $T_g$ and the associated critical exponents \cite{bhatt,kawashima}. An important caveat concerns deviations from finite size scaling which can corrupt the values of the critical parameters estimated from data taken only over a narrow range of small to moderate sizes \cite{mari,palassini}.

In Ising Spin Glasses (ISGs) a basic property which is measured numerically is the mean autocorrelation or memory function  averaged over spins $S_i$ as a function of time $t$ :
\begin{equation}
q(t)=<S_i(t)S_i(0)>
\end{equation}
The freezing temperature $T_g$ is defined as the temperature below which $q(t)$ remains non-zero to infinite $t$ in the thermodynamic limit \cite{edwards}. It can be estimated from the divergence of the relaxation time for very large samples \cite{ogielski}. Alternatively, the standard finite size scaling technique for determining $T_g$ from measurments on moderate size samples is through the dimensionless Binder parameter ratio of the moments of the equilibrium distribution of $q(t)$ \cite{bhatt}:
\begin{equation}
g_L=1/2[3-<q^4>/<q^2>^2]
\end{equation}
The curves $g_L(T)$ for different sizes $L$ should all intersect at $T_g$. Unfortunately in $3d$ ISGs intersections need very high quality statistics to be visible at all \cite{kawashima,ballesteros}and can be subject to severe deviations from finite size scaling \cite{mari2,ballesteros}.

An alternative approach which has been used to identify $T_g$ and associated critical exponents \cite{bernardi}involves the combination of measurements of  three parameters.

First, equilibrium measurements of the spin glass susceptibility
\begin{equation}
\chi_{SG} = L^d[<q^2>]
\end{equation}
on samples of different sizes $L$ obey the scaling rule
\begin{equation}
\chi_{SG}/L^2 \sim L^{-\eta} (1 + const * L^{-\omega}+ …)
\end{equation}
at $T_g$ with $\eta$ being the standard static critical exponent \cite{bhatt} and $\omega$ being the leading static correction to finite size scaling exponent. Below $T_g$ it turns out that the same functional form can be taken to hold with a temperature dependent effective value  $\eta(T)$. In contrast to the case of the two parameters that we will discuss next, direct $\chi_{SG}$ measurements require long preparatory anneals or the use of parallel tempering methods to assure true equilibrium and so are much more demanding numerically than measurements on non-equilibrium parameters. The need to attain true thermal equilibrium in finite computing time means that the maximum sample size for the simulations is in practice limited to $L \sim 20$. Fits to high quality data (including results at small $L$) and allowing for the correction to scaling factor can nevertheless provide accurate values for $\eta(T)$ together with an estimate for the exponent $\omega$ \cite{mari}.

Secondly, following a preliminary anneal at a temperature $T$ near $T_g$ over a long waiting time $t_w$, the initial relaxation of the autocorrelation function takes the form
\begin{equation}
q(t-t_w) = \lambda (t-t_w)^{-x(T)}
\end{equation}
As long as the anneal time $t_w$ is much longer than the measuring run time $t-t_w$, the value of $x(T)$ observed is equal to the equilibrium value (corresponding to infinite $t_w$), and the value $x(T)$ which can readily be obtained on reasonably large sized samples is essentially equal to the infinite size limit \cite{ogielski,rieger}. Direct checks have been made of the independence of the measured $x(T)$ on sample size, and the absence of corrections to finite size scaling beyond a few Monte Carlo Steps per spin (MCS)\cite{rieger}. Hence precise values of $x(T)$, for all intents and purposes equal to the limiting  equilibrium values in the thermodynamic limit, can be obtained by averaging over data on a sufficient number of large samples without the need to anneal to true equilibrium. At $T=T_g$, $x(T_g)$ is related to the standard static and dynamic critical exponents $\eta$ and $z$ through \cite{ogielski}
\begin{equation}
x(T_g)=(d-2+\eta)/2z
\end{equation}

Finally, the out of equilibrium time dependent spin glass susceptibility for  a sample quenched from infinite temperature and annealed towards equilibrium at temperature $T$ is defined by
\begin{equation}
\chi_{SG}^{'}(T)=[<S_i^a(t)S_i^b(t)>^2]
\end{equation}
where $a$ and $b$ are two independent replicas of the same sample initially at infinite temperature at $t=0$. Then for times greater than some microscopic time
\begin {equation}
\chi_{SG}^{'}(t) = At^h(T)(1+B(t^{-w/z}+ higher order)
\end{equation}
where $h(T)$ is the "Huse exponent" \cite{huse} and $w$ is the leading dynamic correction to scaling exponent \cite{parisi}(which in principle does not have to be equal to the leading static correction exponent $\omega$ \cite{ma}). The corrections at short times are the analogue of finite size corrections, as the clusters of correlated spins that are gradually building up with increasing annealing time have a $t$ dependent finite size, even if the sample size $L$ can treated as effectively infinite. The factors $A$ and $B$ may vary with $T$.
At $T=T_g$ \cite{huse}
\begin{equation}
h(T_g)=(2-\eta)/z
\end{equation}
Here, by definition there is no equilibration anneal before a measuring run to determine $h(T)$, and  large samples can readily be used, so it is relatively easy to ensure that for the time range used the system is far from the saturation limit for that particular size, see \cite{zheng}. (Direct tests can be made by comparing data taken for different large values of $L$; when the $h(t)$ values become $L$ independent then one is in the infinite size limit). By averaging over a sufficient number of runs on independent samples, accurate results for of $<\chi_{SG}^{'}(t)>$ can be obtained, so $h(T)$  can be estimated to good precision from a fit including the leading correction to scaling.  Figure 1 shows examples of a plot of data with a fit for the binomial interaction distribution. It turns out that the correction can be measured but is much weaker in the other two cases.

\begin{figure}
\includegraphics[width=9cm,height =6cm,angle=0]{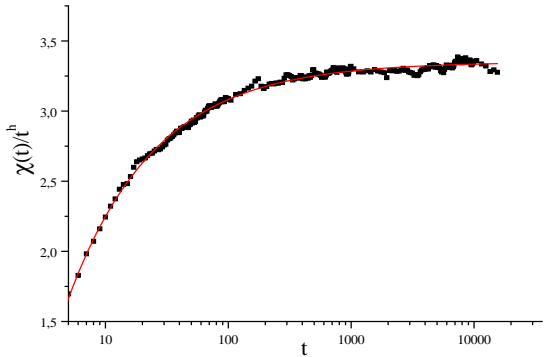} 
 \caption{\label{Figure:1}The non-equilibrium spin glass susceptibility $ \chi_{SG}^{'}(t) $
as a function of time $t$ for the binomial ISG at temperature $T= 1.2$.
The fit is to equation $8$ with the Huse exponent $h(T)= 0.398$ and the
leading correction exponent $w/z= 0.61$.}
\end{figure}



Suppose we now make measurements of the three independent parameters $x(T_i),  h(T_i)$ and $\eta(T_i)$ at a series of test temperatures $T_i$ around $T_g$. We can define a new effective parameter $h^{*}(T_i)$ through
\begin{equation}
h^{*}(T_i)=2x(T_i)(2-\eta(T_i))/(1+\eta(T_i))
\end{equation}
and we can plot both $h(T_i)$ measured directly and the derived $h^{*}(T_i)$ against $x(T_i)$, with $T_i$ as an implicit parameter.

For one given system, because of the definitions of $h$ and $h^{*}$ at $T_g$ in equations $6, 9$ and $10$, consistency dictates that the curves for $h(x)$ and for $h^{*}(x)$ must intersect at a point which corresponds to $T_i = T_g$ and which therefore lies at the point $x=x(T_g)$ and $h =h^{*}=h(T_g)$.
If standard universality holds, the values of the static critical exponents $\eta(T_g)$ and of the dynamic critical exponents $z(T_g)$ at the appropriate critical temperatures should be identical for all short range interaction ISGs in a given dimension. The form of the random interaction distribution should not be pertinent. Therefore the curves $h(x)$ and $h^{*}(x)$ for {\it all} ISGs in a particular dimension should intersect at one single point corresponding to the universal $[h(T_g),x(T_g)]$.

We have made measurements on standard three dimensional ISGs on  simple cubic lattices with toroidal boundary conditions. The three systems chosen have random near neighbour interaction distributions which are binomial, Gaussian, and decreasing exponential respectively.  The data for $x(T)$ and $h(T)$ are from our own simulations. To obtain these parameters samples of maximum size $L = 28$ were studied, with an average being taken over $6600$ independent samples. For the $x(T)$ measurements preliminary anneals were performed to at least $10^6$ MCS, and the run times used were up to $10^4$ MCS. For the binomial interaction distribution and for the Gaussian interaction distribution, $x(T)$ values are in excellent agreement with data shown in \cite{ogielski} and \cite{rieger} at all temperatures where direct comparisons can be made. There do not appear to be published measurements for the decreasing exponential distribution. For the $h(T)$ measurements, runs were made to $10^4$ MCS. Our binomial interaction data are in good agreement with but are more accurate than those of \cite{huse,bray}. We are not aware of published $h(T)$ data on the other systems. For the $\eta(T)$ values we rely on accurate size dependent spin glass susceptibilty $\chi_{SG}$ results by \cite{kawashima,cruz} and \cite{marinari} for the binomial distribution up to $L=24$ and the Gaussian distribution up to $L=16$ respectively. The $\eta(T)$ data for the decreasing exponential distribution correspond to our own measurements of $\chi(T)$ up to only $L=8$. The statistical precision of the $x(T)$ and $h(T)$ values is about $\pm 0.005$ for all three systems, and as we have explained the residual systematic error due to incorrect extrapolation to infinite size can be taken to be small. The precision of the $\eta(T)$ values is about $\pm 0.02$ for the binomial and Gaussian interaction systems, with the uncertainty being mainly statistical with a  possible contribution from incomplete equilibration. The uncertainty is greater for the decreasing exponential case because of relaxation rates for equilibration which are even slower than for the other interactions, limiting the sizes $L$ which we could anneal to full equilibrium.

The results are presented in Figure 2. For each of the three systems individually, the critical values $x(T_g)$ and $h(T_g)$  can be estimated directly from the $[h(x), h^{*}(x)]$ intersection point. The corresponding critical values $\eta$ and $z$ follow immediately from equations $4$ and $6$. As we know $x(T)$, the values of the critical temperatures $T_g$ follow also from the positions of the intersection points. The values estimated are given in Table 1.  The values of the critical temperatures can be compared to other published estimates \cite{ogielski,bhatt,kawashima,palassini,marinari,ballesteros}.

\begin{figure}[h]
\includegraphics[width=9cm,height =6cm,angle=0]{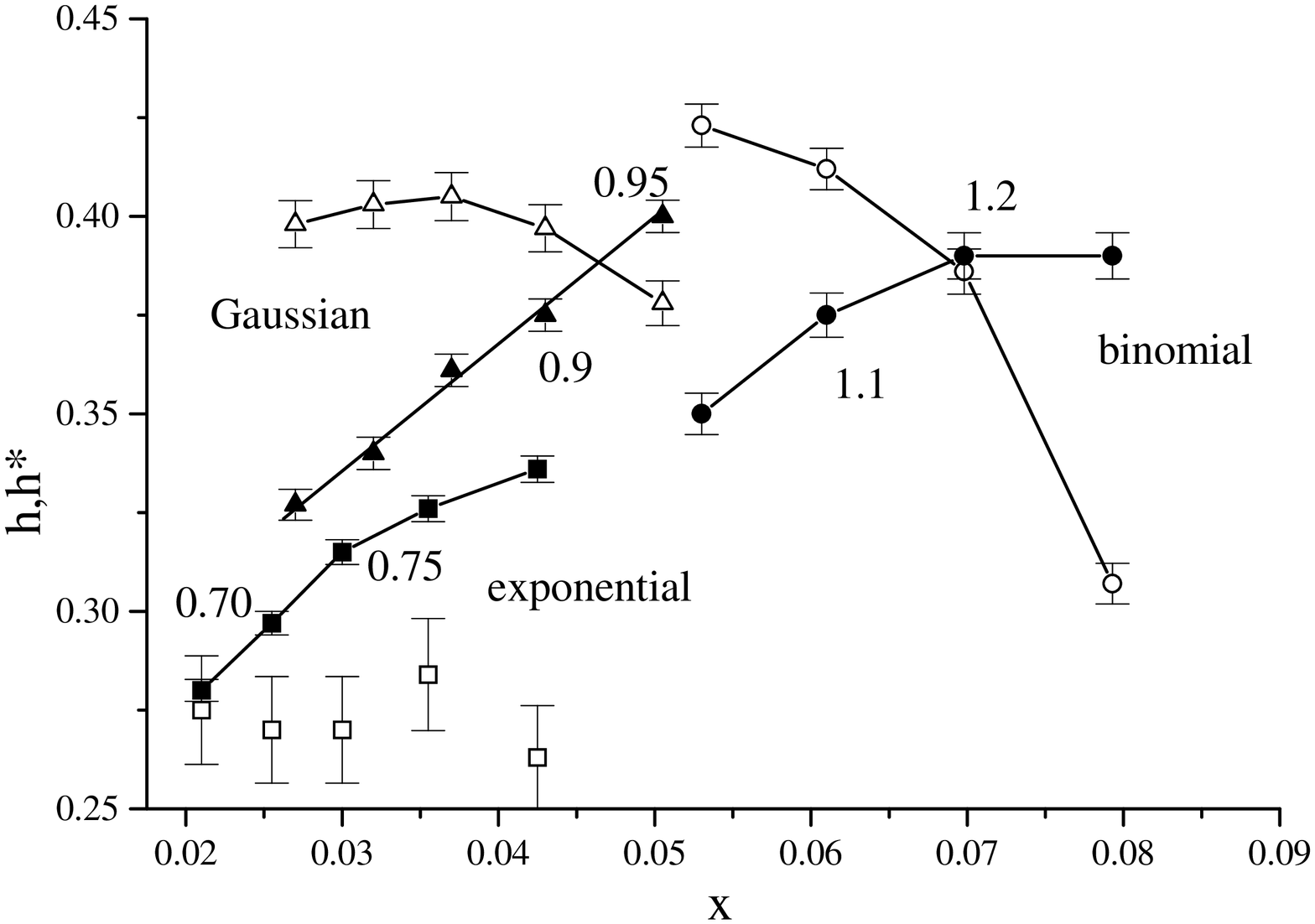} 
 \caption{\label{Figure:2}The effective exponents $h(x)$ (closed symbols) and $h^{*}(x)$ (open symbols) as functions of the exponent $x$, with the temperature $T$ as an implicit parameter.  Circles : binomial distribution, triangles : Gaussian distribution, squares : decreasing exponential distribution. The numbers indicate temperatures for the different distributions. The intersection points of $h(x)$ and $h^{*}(x)$ correspond to the critical temperature $T_g$ for each distribution.}
\end{figure}

A cursory glance at  Figure 2 shows that the intersections of the $h(x)$ and $h{*}(x)$ curves for the three different systems do not coincide at one single point in the $[h,x]$ plane, so the naive universality rule, which would have the ISG transitions lying within a single universality class whatever the form of the interaction distribution, is violated.

If we consider the tabulated values, it can be seen that the various exponents do not vary in a random way. Going from one system to the next, as the kurtosis of the interaction distribution increases and the $T_g$ values diminish, $\eta$ becomes more negative and $z$ increases. This is just the tendency one would expect if the exponents were tending regularly towards the $T_g=0$ values, which are $z=\infty$, and $\eta=-1$  in dimension $3$. We will discuss elsewhere the behaviour of the correction to scaling exponent $w$ \cite{FSS}, but we can note that the analyses indicate high values of $w$.  In \cite{mari} a high $\omega$ estimate was linked to the fact that dimension $3$ is far below the upper critical dimension, $6$. High values of $w$ and $\omega$ are favourable, as they mean that corrections to finite size scaling die out relatively rapidly with $t$ and $L$ respectively.

\begin{table}[h]
\caption{\label{Table:1} Estimates of $T_g$, $\eta$ , $z$ and $w/z$ for the 3d ISG
with different interaction distributions }
\begin{ruledtabular}
\begin{tabular}{ccccc}
Interaction         & $T_g$         & $\eta$            & $z$       & $w/z$ \\

Binomial        & $1.190(15)$   &  $-0.20(2)$   & $5.85(10)$    &  $0.48(3)$   \\
Gaussian           &  $0.920(15)$    &  $-0.42(2)$   & $6.45(10)$ &   $ 0.62(3)$      \\
Decreasing Exponential  &  $0.70(5)$    &  $ -0.54(5)$ & $8.8(3)$&   $0.95(2)$     \\

\end{tabular}
\end{ruledtabular}
\end{table}

Universality has a strong aesthetic appeal, and the general physical argument outlined in the introduction would seem very robust ; nevertheless in a different context there are examples of systems with many attractors where critical exponents vary continuously as a function of a control parameter \cite{noh}. Evidence has been found for universality in Migdal-Kadanoff (MK) spin glasses \cite{brazil}; however MK spin glasses are known to have a "droplet" structure \cite{Mkdroplet}, while there are strong arguments in favour of a more complex structure in finite dimension ISGs (see for instance \cite{marinari2}). In the finite dimension spin glass context the $\epsilon$ expansion below the upper critical dimension in ISGs has proved totally intractable \cite{dedominicis}. Parisi et al \cite{parisi2} state that the classical tools of RG analysis are not suitable for spin glasses. They introduce a non-standard coarse graining technique that is performed on the overlap probability measure. Within this generic approach a much richer structure of universality classes could appear in spin glasses, and in particular the form of the interaction distribution could play a fundamental role. Thus in ISGs there could well exist  fundamental reasons to expect changes of critical exponents with interaction distributions like those we observe empirically.

We would like to acknowledge an allocation of computing time from the French national computing center IDRIS, and very useful conversations with Ludovic Bertier and with Francesco Rosati.

\end{document}